\documentclass{aa}

\usepackage{natbib}
\bibpunct{(}{)}{;}{a}{}{,}
\usepackage{graphicx}
\usepackage{amsmath,amssymb}
\usepackage[dvipdfm,breaklinks=true,bookmarks=true,hypertexnames=false,
bookmarksnumbered=true,bookmarksopen=false,bookmarks=true,bookmarksnumbered=true]{hyperref}

\newcommand{\kms}{\, {\rm km\, s}^{-1}}

\newcommand{\kpc}{\, {\rm kpc}}

\newcommand{\Mpc}{\, {\rm Mpc}}

\newcommand{\der}{{\rm d}}
\newcommand{\reffig}[1]{Fig.~\ref{#1}}

\newcommand{\scrit}{\Sigma_{\rm crit}}

\newcommand{\be}{\begin{equation}}
\newcommand{\ee}{\end{equation}}
\newcommand{\ba}{\begin{eqnarray}}
\newcommand{\ea}{\end{eqnarray}}
\newcommand{\brr}{\begin{array}}

\newcommand{\err}{\end{array}}
\begin{document}

\title{Probing dark matter caustics with weak lensing}
\author{Raphael Gavazzi \inst{1,3} \and Roya Mohayaee \inst{2} \and Bernard Fort \inst{2}}

\offprints{R. Gavazzi, \email{rgavazzi@ast.obs-mip.fr}}
\date{}
\institute{Laboratoire d'Astrophysique, UMR5572 CNRS \& Universit\'e Paul Sabatier, 14 Av Edouard Belin, F-31400 Toulouse, France
  \and Institut d'Astrophysique de Paris, UMR7095 CNRS \& Universit\'e Pierre \& Marie Curie, 98bis Bd Arago, F-75014 Paris, France \and
  Oxford University, Astrophysics,
  Denys Wilkinson Building, Keble Road, Oxford OX1 3RH, UK
}

\abstract{
Caustics are high-density structures that form in collisionless media.
Under self-gravity, cold dark matter flows focus onto caustics 
which are yet to be resolved in numerical simulations and observed 
in the real world. If detected, caustics would provide
strong evidence for dark matter and would rule out alternative models
such as those with modified dynamics. Here, we demonstrate how they 
might be observed in weak lensing data. We evaluate the
shear distortion and show that its radial profile is marked by a
characteristic sawtooth pattern due to the caustics in dark matter haloes
that form by selfsimilar accretion. We discuss 
the observational complications, mainly due to the poor knowledge 
of the virial radii of the haloes and demonstrate that a superposition 
of about 200 cluster-size
haloes would give a signal-to-noise ratio which is sufficiently large 
for the detection of caustics with ground-based observations. This 
number is reduced to 60 for space-based observations.
These bounds can be easily achieved by the ongoing wide field optical
surveys such as {\small CFHTLS} and the future space-based projects 
{\small SNAP} and {\small DUNE} which have to be accompanied
by an X-ray follow-up of the selected clusters
for a precise determination of their virial radius.
}
\maketitle
%======================================================================

%======================================================================
\section{Introduction}
The nature of dark matter, which constitutes about $30\%$ of the mass of the
Universe, remains largely unknown. Results from cosmic microwave background
explorers and large-scale galaxy surveys suggest that dark matter is cold with
little velocity dispersion \citep[{\it e.g.}~][]{spergel03,tegmark04}. 
If so, then its evolution is mainly governed by its self-gravity 
and expressed by Jeans-Vlassov-Poisson equations.
The collisionless nature of dark matter predicts the formation of multi-stream
regions bounded by very high density manifolds known as {\it caustics}.

In a cold dark matter Universe, caustics can form on large scales of
many megaparsecs, manifesting in the filamentary structure of galaxy
distribution \citep{shandarin89} and also at smaller scales 
of a few parsecs or kiloparsecs in dark matter haloes. This is well
illustrated in some pionering numerical works \citep{melott89}.
Due to their abundance, the rich observational and
numerical data and their high density contrasts, 
haloes are likely areas for the caustics.

Analytic models for the formation and evolution of dark matter haloes 
are still rare and most works are based on the selfsimilar 
accretion model \citep[]{gott75,gunn77,fillmore84,bertschinger85}.
In this model, first proposed to explain the rotation curve of galaxies,
haloes form by temporally self-similar collapse of dark matter shells onto
an initially over-dense perturbation.
Dark matter shells initially expand until they reach
their turnaround radii where they separate from the
background expansion and collapse. After collapse they re-bounce and collapse
again and the density profile settles asymptotically into a power-law
which is convolved with singular spikes, namely with caustics.

In the following we shall refer to such caustics as "outer caustics".
They are suggestive of the sharp stellar shells observed around
giant ellipticals which can arise in the
merger of galaxies \citep{malin80,quinn84,fort86}.
The main observational difference between the merger and selfsimilar spherical
infall is that the former predicts that the caustics are interleaved with the
caustic radii alternating on opposite sides of the galaxy and
the latter predicts concentric spherical shells.

The spherically symmetric model has been extended to consider
infalling matter with angular momentum and calculated the properties
of an additional kind of "inner caustics" with torus-like topology
\citep{sikivie98,sikivie99}. Such accretion with angular momentum is
more relevant for galaxy-size haloes than for clusters of galaxies dominated
by radial infall.

Here, we focus on the outer caustics of cluster-size haloes and
argue that they can be reasonably approximated by selfsimilar infall models.
In the original version of this model, dark matter is {\it absolutely} cold,
{\it i.e.} with zero velocity dispersion, and caustics are infinitely
thin concentric spherical shells with diverging densities.
However, realistically, dark matter has a small velocity dispersion
and these shells have finite thicknesses. The thicknesses of the caustics
would however remain very small (due to the coldness of dark matter)
and thus they would contain very little mass in spite of their
significant density. Various characteristics of the caustics such as
their density profile, their thickness and their approximate maximum density 
for a low velocity dispersion dark matter medium have been recently evaluated
\citep[][hereafter MS05]{mohayaee05}.

Detection of dark matter caustics remains a challenging problem both
for 3-dimensional numerical simulations and for observations. 
Dark matter annihilation in the caustics has already been studied 
\citep[See {\it e.g.}~][]{sikivie97,hogan01,mohayaee05,pieri05}. Since 
the flux of the annihilation products, {\it e.g.}~ $\gamma$-rays
or antiproton flux, depends on the square of the local
density\footnote{See {\it e.g.}~ \citet{donato04} for the antiproton flux 
and \citet{bertone05} for a general recent review.}, caustics with their 
sharply-peaked densities would be the likely places for significant
emissivity. Such kind of observations are promising for the eventual detection
of dark matter in the caustics. However, the major problem with dark matter 
detection through gamma-ray emission is the severe background 
contamination of the signal and the low signal-to-noise ratio of
present day observations.

It has been shown that rotation curves of galaxies might be sensitive
to the presence of inner caustics and claimed a marginal detection based
on an ensemble average over 32 rotation curves \citep{kinney00}. Inner
caustic rings are located in the plane of the disk and are likely to
modify rotation curves more efficiently than outer caustics with
spherical symmetry. Moreover, rotation curves only probe the dark
matter potential of galactic haloes where a large amount of tracers
(gas or stars) is available. Therefore, outer caustics which are
located well beyond the observable tracers cannot leave a detectable
imprint on rotation curves.

Gravitational lensing provides a promising alternative tool.
Since lensing probes the projected density profile with no regards to
the nature or dynamical state of the deflecting mass, it should be
sensitive to the caustics. The lensing properties of dark matter caustics
and particularly their efficiency in magnifying and/or producing multiple
images of background sources has already been investigated \citep{charmousis03}.\\
Inner caustics may be dense enough to produce substantial magnification
and small separation multiple images as in {\it micro-lensing} by compact objects.
High magnification events due to such caustics may explain the anomaly in flux
ratios observed in multiply-imaged quasars which are hardly reproduced by a smooth
halo or even subhalos \citep[see {\it e.g}~][]{dalal02,kochanek04}.\\
Conversely, it has been found that outer caustics are inefficient in
magnifying distant sources and would yield at most a few percent net
magnification or shear \citep{charmousis03}. As we shall detail below,
the selfsimilarity of dark matter accretion implies that outer
caustics occur at the same radius provided physical radii are properly
rescaled by the halo mass. Hence the tiny lensing signal of caustics could show up
statistically by averaging over many rescaled haloes.
Such a statistical approch is much more complicate for the lensing-based
detection of inner caustic rings because of the random orientation of rings
(or angular momenta).

In this work, we consider the lensing properties of the outer caustics
only and we propose the {\it weak-lensing} effect as a potential 
way to detect caustics. We demonstrate that the caustics will produce sharp
variations in the projected surface mass density around haloes. Depending
on the height and width of caustics, gravitationally-distorted background
galaxies will experience local variations of shear.

If the aforementioned universal property of haloes is fullfilled, 
then the rapid progress in X-ray and
lensing observations of cluster of galaxies may offer one possibility to
observe dark matter caustics. Deep and wide surveys such as CFHTLS
will cover fields of view of a few hundreed square degrees and will
provide us with a useful material for the detection of dark matter caustics
through the capability of mass to coherently stretch the image of the
background galaxies. They will provide us with a large enough number
of galaxy clusters\footnote{$\sim 5$ per square degree,
  \citep{hennawi05}} to achieve the required level of signal-to-noise
ratio. Wide field spatial surveys will be
even more powerful for the investigation of the lensing properties of
dark matter haloes and their associated caustics. An important
requirement for the detection of caustics is the measurement of virial
radii of the clusters which can be provided by X-rays
 \citep[{\it  e.g.}~][]{arnaud05}. 

Throughout this paper, we assume an Einstein-de Sitter Universe but our
results should be qualitatively similar in a concordance $\Lambda$CDM model.
The role of dark energy becomes important at low redshifts ($\sim
0.2$) which we expect to occur well after the formation of the typical dark
matter haloes we consider here. Furthermore, once a particle turns around
and collapses, it separates from the background expansion and its subsequent
motion should not be affected by the $\Lambda$ term.

The paper is organized as follows. We review the three-dimensional
and projected properties of self-similar haloes and caustics
in Sect. \ref{sec:caustics}. We derive the lensing signal for a single halo,
compare it to the noise level of fiducial observations and estimate
the number of haloes required to achieve a significant signal-to-noise
ratio in Sect. \ref{sec:lensing}, where we also examine the ability of
weak lensing to constrain the velocity dispersion of dark matter particles.
We summarize, discuss the prospectives for future works 
and conclude in Sect. \ref{sec:conclu}.

%======================================================================
\section{Dark matter caustics}\label{sec:caustics}
%------------------------------------
\subsection{Tri-dimensional key equations}

In an Einstein-de Sitter Universe a spherical overdensity
expands and then turns around to collapse. After collapse and at late
times, the fluid motion becomes selfsimilar: its form remains
unchanged when its length is re-scaled in terms of the radius of the
shell that is currently at the ``turn around'' and is falling onto the
galaxy for the first time. Physically, selfsimilarity arises because
gravity is scale-free and because mass shells outside the initial
overdensity are also bound and turn around at successively later
times. Self-similar solutions give power-law density profiles
whose exact scaling properties depend on the central boundary
conditions and on whether the fluid is collisionless or collisional
\citep{gott75,gunn77,fillmore84,bertschinger85}. The density profile
obeys a power-law on the scale of the halo which provides an explanation
of the flattening of the rotation curves of the galaxies.
However, on smaller scales the density profile contains many spikes
({\it i.e.} caustics) of infinite density. The position and the time
of formation of these caustics are among the many properties that have
been studied in the selfsimilar infall model \citep{bertschinger85}.

In the presence of a small velocity dispersion the maximum density and
thickness of the caustic shells and their density profiles have been
evaluated in the framework of a selfsimilar collapse model (MS05). 

The global halo density profile,
asymptotically reached in this process, is well-approximated by
\begin{equation}\label{eq:rho3Dhalo}
  \frac{\rho_{\rm halo}(\lambda)}{\rho_H} 
\sim \frac{2.8\,\lambda^{-9/4}}{(1+\lambda^{3/4})^2}\,,
\end{equation}
where  $\lambda=r/r_{ta}$ with $r_{ta}$ the present turnaround radius
of the halo and $\rho_H=3H^2/8\pi G$ the background density. The turnaround
radius can be easily computed using the virial radius. Within the virial radius,
$r_{\rm vir}=r_{200}$, the mean density, $\rho_{\rm vir}$, is, by definition,
$200$ times the background density. Thus, using the density profile
\eqref{eq:rho3Dhalo}, we obtain the following relationship between the
turnaround and the virial radii: $r_{\rm ta}\sim\, 4\,r_{\rm vir}$.

For a perfectly cold dark matter medium, the density profile close
to a caustic at $\lambda_k $ is \citep{bertschinger85}
\begin{equation}\label{eq:rho3Dsing}
  \frac{\rho_0(\lambda)}{\rho_H}= \frac{G_k}{\sqrt{\lambda_k-\lambda}}\,;
\qquad\qquad \sigma=0.
\end{equation}
with
\begin{equation}
G_k=\frac{\pi^2}{4\sqrt{-2\lambda_k''\,}}\,\frac{e^{-2 \xi_k/3}}{\lambda_k^2 } \;,
\label{Gk}
\end{equation}
where the values of the various quantities $\xi_k$, $\lambda_k$, $\lambda_k''$
[and $\Lambda_k$ which will appear in the coming expression
(\ref{eq:thickness})] are given in Table \ref{table:caustic-parameters}
[see MS05 for a detailed description of these parameters].

%_______________________________________________________________
\begin{table*}
\centering
% \begin{minipage}{180mm}
%  \begin{tabular}{@{}llrrrrlrlr@{}}
\begin{tabular}{@{}cccccccc@{}}\hline\hline
k\,\,\,\,\,\, &  $\xi_k$\,\,\,\,\,\,   &  $\lambda_k$\,\,\,\,\,\, &
$\left(d^2\lambda/d\xi^2\right)_k$ & $\Lambda_k$  & \qquad $\Delta r_k$ (pc) 
& \qquad $\rho_{\rm max}/\rho_H/(\sqrt{r_{\rm vir}})\,\, pc^{-1/2}$ & $\rho_{\rm halo}/\rho_H$\\
\hline
1     &    0.985    &    0.368    &       -5.86     &   -0.0704     & $5.6\, 10^{-4}$    &  233    &  12   \\
2     &    1.46     &    0.237    &       -11.2     &   -0.0254     & $2.6\, 10^{-4}$    &  431    &  39   \\
3     &    1.76     &    0.179    &       -16.7     &   -0.0135     & $1.7\,10^{-4}$     &  640    &  82   \\
4     &    1.98     &    0.146    &       -22.3     &   -0.00854    & $1.2\,10^{-4}$     &  850   &  139  \\
5     &    2.16     &    0.124    &       -28.0     &   -0.00591    & $9.1\, 10^{-5}$    &  1067   &  210  \\
6     &    2.31     &    0.108    &       -33.9     &   -0.00437    & $7.3\, 10^{-5}$    &  1290   &  297  \\
7     &    2.43     &    0.096    &       -39.8     &   -0.00337    & $6.0\,10^{-5}$     &  1532   &  398  \\
8     &    2.55     &    0.087    &       -45.7     &   -0.00266    & $5.1\, 10^{-5}$    &  1749   &  513  \\
9     &    2.64     &    0.079    &       -51.7     &   -0.00221    & $4.4\, 10^{-5}$    &  2010   &  640  \\
10    &    2.73     &    0.073    &       -57.8     &   -0.00182    & $3.8\, 10^{-5}$    &  2253   &  785  \\
\hline\hline
\label{table:caustic-parameters}
\end{tabular}
\caption{\footnotesize The non-dimensional times ($\xi_k$), positions ($\lambda_k$), 
thicknesses ($\Delta r_k$), maximum densities ($\rho_{\rm max}$)
of the first ten caustics, the halo density itself ($\rho_{\rm halo}$)
at the positions of these caustics and other parameters used in expressions
(\ref{Gk}), (\ref{eq:rho3Ddef}), (\ref{eq:thickness}) and (\ref{eq:maxdens}). 
The present velocity dispersion is that  
for neutralinos which is about $0.03$ cm/s. It is worth emphasizing 
that the thicknesses of the caustics
do not depend on the halo parameters but only on the velocity
dispersion of dark matter [see expression (\ref{eq:thickness})].
It is instructive to compare the last two columns, which show that for a
cluster-size halo with $r_{\rm vir}\sim 1$ Mpc, the caustic density can be many orders of
magnitude higher than the halo density itself.}
%\end{minipage}
\end{table*}
%----------------------------------------------------------------------------

When the temperature of particles is not strictly zero and the velocity of
particles is distributed according to the distribution function $f(v)$,
caustic positions are shifted by a small value $\delta\lambda$ and
the caustic density is modified as (MS05):
\be
  \rho_\sigma(\lambda)= \int \der v\, \rho_0\,[\lambda - \delta\lambda(v)]\,
f(v)\,\,;
\qquad \sigma\not=0 \qquad 
\ee
In this work, we choose a top-hat velocity distribution function (MS05).
Then, the density close to the $k$-th caustic in the halo is given by 
\begin{equation}\label{eq:rho3Ddef}
\frac{\rho_\sigma(\lambda)}{\rho_H} = \frac{G_k}{\Delta_k} \left\{
\begin{array}{ll}
\sqrt{\lambda_k^+-\lambda} - \sqrt{\lambda_k^--\lambda} & 
\text{~~for~}\lambda< \lambda_k^-\;,\\
\sqrt{\lambda_k^+-\lambda} & \text{~~for~}\lambda_k^-<\lambda<\lambda_k^+\;, \\
0 &\text{~~for~}\lambda> \lambda_k^+\;,
\end{array}\right.
\end{equation}
where $\lambda_k^-=\lambda_k-\Delta_k$ and $\lambda_k^+=\lambda_k+\Delta_k$
and the thickness of the $k$-th caustic, $\Delta_k$, 
in nondimensional coordinate is given by
\be
r_{\rm ita}\, \Delta_k=\Delta r_k={(3\pi)^{2/3}\over 4} 
e^{5\xi_k/9}\,\vert\Lambda_k\vert\,\,
 t\,\sigma(t)
\label{eq:thickness}
\ee
and $r_{\rm ita}$ is the initial turnaround radius and $\sigma(t)$
is the value of the velocity dispersion of dark matter particles at time $t$
which is that at decoupling re-scaled by the expansion
factor\footnote{Hereafter, velocity dispersion is given at the present time,
$z=0$. For instance, neutralinos have $\sigma\sim 0.03$ cm/s and
 axions have $\sigma\sim 10^{-7}$ cm/s.}.
The standard spherical collapse model yields a relation
$r_{\rm ita}=2 r_{\rm vir}$ for a constant overdensity
[see {\it e.g.}~ chapter 5.10 of \citep{padmanabhan02b3}], However, the
real value of the initial turnaround radius would be lower due to the
continuous accretion by the halo. 
Thus the physical thickness of
the caustic, $\Delta r_k$, depends only on its position in the halo
and the nature of dark matter. 

The maximum density at each caustic position can also be approximated
by the expression (see Table \ref{table:caustic-parameters} and also MS05)
\be\label{eq:maxdens}
\rho_{\rm max}={2\,G_k e^{-5\xi_k/18}\over (3\pi)^{1/3}\sqrt{-\Lambda_k\,}}\,\,
\sqrt{r_{\rm ta}\over t\,\,\sigma(t)}\,\, \rho_H
\ee

Expressions \eqref{eq:rho3Ddef}, \eqref{eq:thickness} and
\eqref{eq:maxdens} provide us with a sufficient basis for the evaluation
of magnification and shear due to dark matter caustics. Although in
the rest of this work we directly integrate expression \eqref{eq:rho3Ddef}
and never use \eqref{eq:maxdens}, values for the latter are given
in Table \ref{table:caustic-parameters} to demonstrate the density contrast
of each caustic with respect to its host halo.

Although the selfsimilar model might seem naive, it provides a good
approximation for outer caustics in galaxy-cluster haloes which are
not significantly disrupted by merger and substructures and to a good
approximation are spherical. Furthermore, fluctuations caused by
large scale structure would already be averaged out in the statistical
evaluation of the shear.

%------------------------------------
\subsection{Projected densities}\label{subsec:projec-dens}
Since we are concerned with the lensing properties of caustics, we have to
calculate the projected density profile for the caustics and for the halo.
The Abel integral relates the 3-dimensional density ($\rho$) 
and the 2-dimensional density profiles ($\Sigma$) by
\begin{equation}
  \Sigma(\lambda)= 2 r_{ta} \int_{\lambda}^\infty \frac{\rho(\lambda') 
\lambda' \der \lambda'}{\sqrt{\lambda'^2-\lambda^2}}\;.
\end{equation}
We numerically integrate the above expression, for the density profiles
\eqref{eq:rho3Dsing} and \eqref{eq:rho3Ddef}. The halo surface mass
density can be evaluated analytically
if we neglect the $(1+\lambda^{3/4})^2$ term in
equation \eqref{eq:rho3Dhalo}, {\it i.e.} at small scales $\lambda\ll 1$, 
yielding the approximate projected halo profile
\begin{equation}
\Sigma_{\rm halo}(\lambda) \simeq  r_{ta} 7.56 \lambda^{-5/4}\rho_H\,.
\end{equation}

Let us define the mean projected density enclosed by the cylinder of
radius $\lambda$:
\begin{equation}
  \overline{\Sigma}(\lambda)= \frac{2}{\lambda^2} 
\int_{0}^\lambda  \Sigma(\lambda') \lambda'\der \lambda'\;.
\end{equation}
Thus, under the hypothesis of small $\lambda$, the mean projected
density for the halo component is
$\overline{\Sigma}_{\rm halo}(\lambda)= \frac{8}{3} \Sigma(\lambda)$.

%...................
\begin{figure*}[htbp]
  \centering
  \resizebox{8.5cm}{!}{\includegraphics{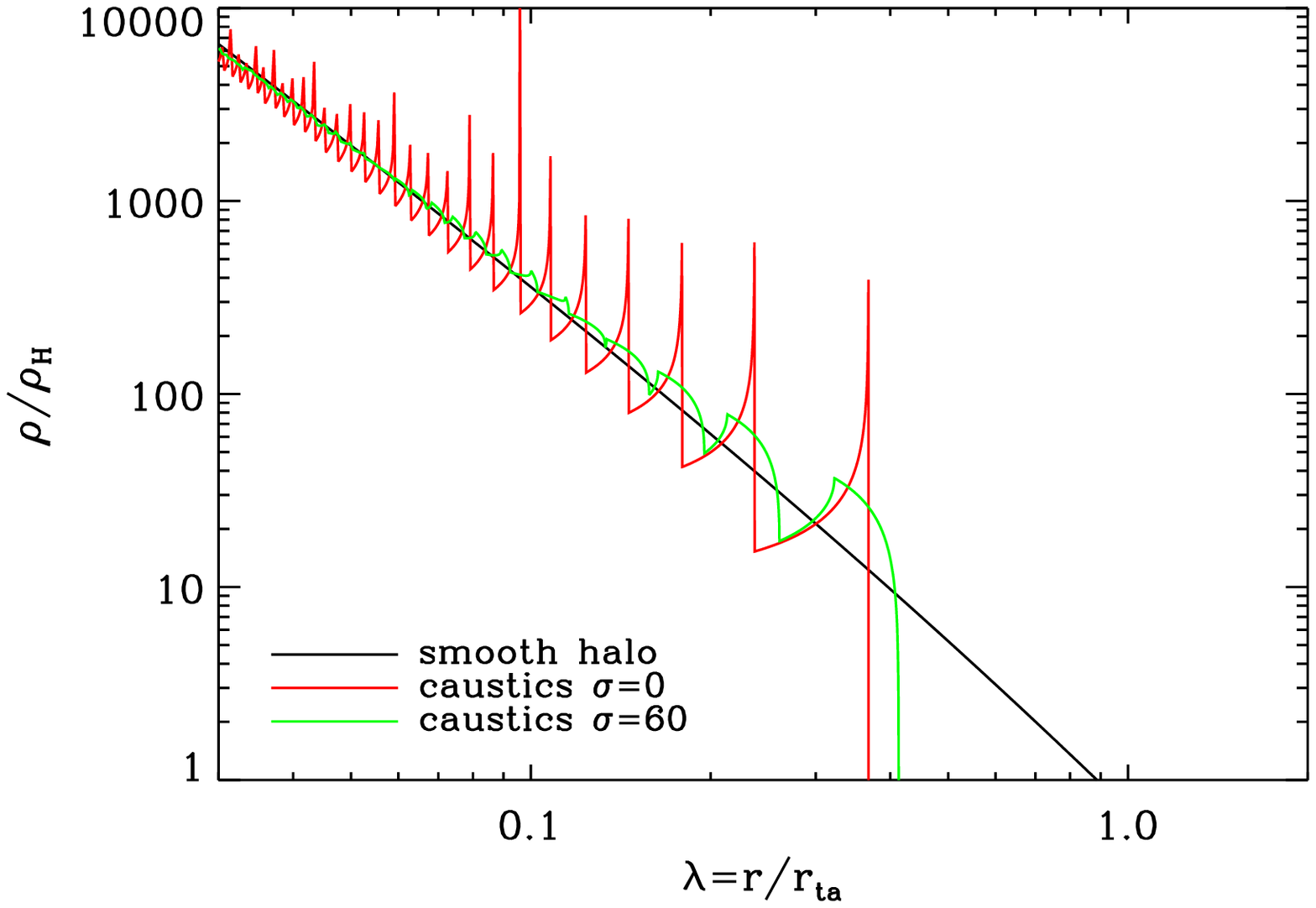}}
  \resizebox{8.5cm}{!}{\includegraphics{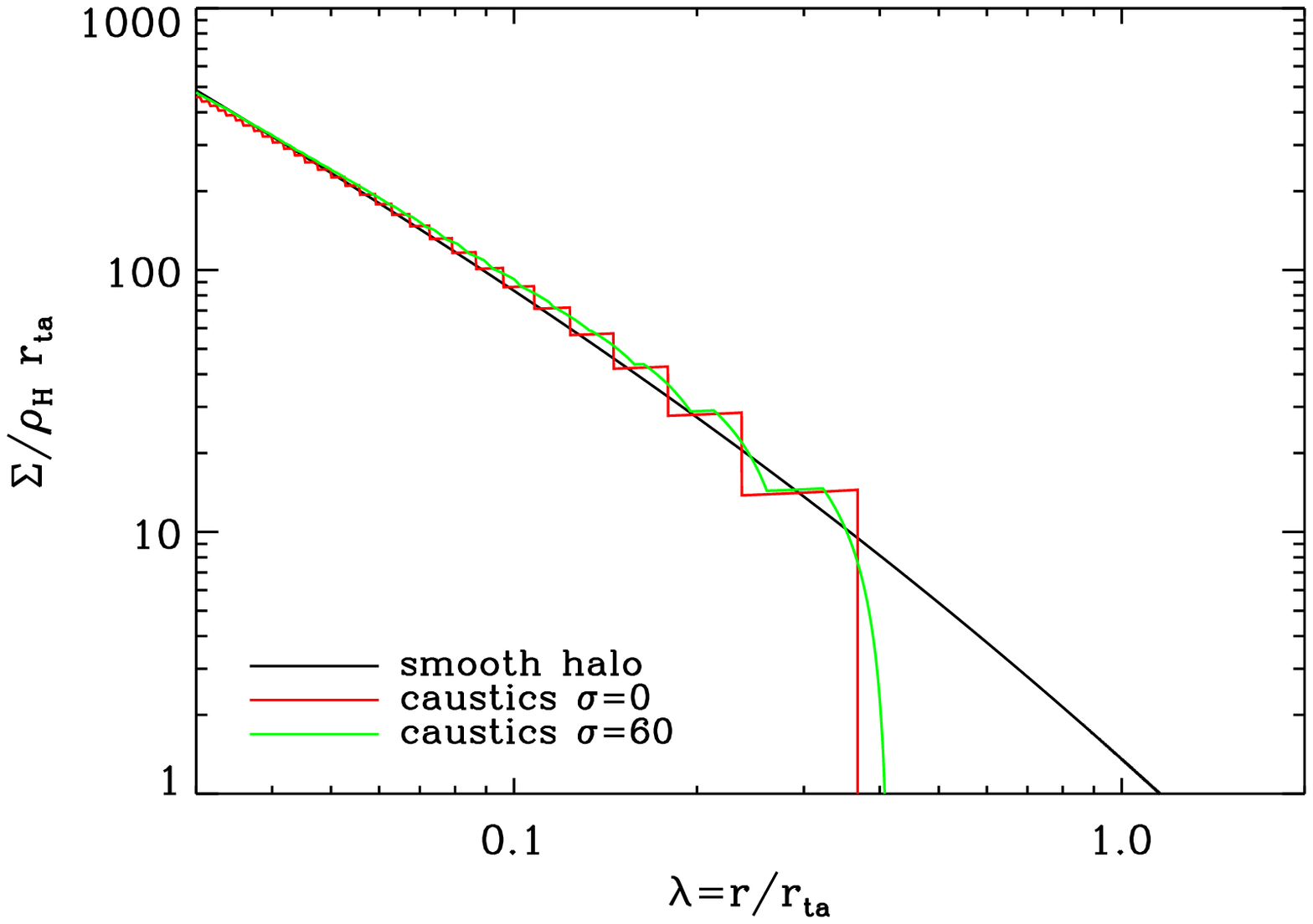}}
  \caption{\footnotesize {\it Left panel: } 
    In units of the background density, $\rho_H$, we show the 3D
    density profile for the halo (black), for the caustics in a
    perfectly cold medium (red) and for the caustics smoothed out by a
    warm/hot dark matter with velocity dispersion $\sigma=60\kms$
    (green). Red spikes are all singular with infinite density and
    limited here to finite values because of finite resolution. \ \ \
    {\it Right panel: } The same colour-coding for the projected 2D
    density profiles in units of $\rho_H r_{ta}$. When projected,
    caustic peaks are smoother and look like a flight of stairs.
    Therefore, we expect low lensing magnifications close to
    the caustics (see Section \ref{sec:lensing}). 
  }
  \label{fig:dens3-2}
\end{figure*}
%...................

Figure \ref{fig:dens3-2} shows the 3D (left) and 2D (right) density
profiles for the halo and the caustics. We consider the case of a
perfectly cold dark matter with very peaked caustics and the case in
which they are smoothed out by a finite velocity dispersion
$\sigma \sim 60 \kms$. This value is very high for most cold dark
matter models but its extremity illustrates well the peak dilution 
and shows that caustics survive a large thermal softening. When
considering the projected density, instead of spikes we rather see
stairs that come from the analytic $1/\sqrt{r_k-r}$ singularity of
caustics. Consequently even with singular caustics, the projected
density profile is not peaked. The implications for lensing are
discussed in the next section.

%======================================================================
\section{Weak lensing}\label{sec:lensing}
%------------------------------------
\subsection{General equations}
The fundamental quantity for gravitational lensing is the lens potential
$\psi(\vec{\theta})$ at the angular position $\vec{\theta}$ which is related
to the surface mass density $\Sigma(\vec{\theta})$ projected into the lens
plane through :
\begin{equation}
\psi(\vec{\theta}) = \frac{4 G}{c^2} \frac{D_{\rm l} D_{\rm s}}{D_{\rm ls}} 
\int \der^2 \theta' \Sigma(\vec{\theta}') \ln\vert \vec{\theta}-\vec{\theta}'\vert\;,
\end{equation}
where $D_{\rm l}$, $D_{\rm s}$ and $D_{\rm ls}$ are angular distances to the lens,
to the source and between the lens and the source respectively. The deflection
angle $\vec{\alpha}=\vec{\nabla}\psi$ relates a point in the source plane
$\vec{\beta}$ to its image(s) in the image plane $\vec{\theta}$ through the
lens equation $\vec{\beta}=\vec{\theta}-\vec{\alpha}(\vec{\theta})$.
The local relation between $\vec{\beta}$ and $\vec{\theta}$ is the
Jacobian matrix $A_{ij}=\partial \beta_i/\partial \theta_j$ :
\begin{equation}\label{eq:jacob}
  A_{ij} = \delta_{ij} - \psi_{,ij} = \left(\begin{array}{cc}
      1-\kappa-\gamma_1 & -\gamma_2 \\
      -\gamma_2  & 1-\kappa+\gamma_1
    \end{array}\right)\;.
\end{equation}
with the convergence $\kappa(\vec{\theta})=\Sigma(\vec{\theta})/\scrit$ 
directly related to the surface mass density via the critical density
\begin{equation}\label{eq:scrit}
  \scrit = \frac{c^2}{4 \pi G}\frac{D_{\rm ls}}{D_{\rm l}D_{\rm s}}\;,
\end{equation}
and the 2-component shear $\gamma=\gamma_1+i \gamma_2$ in complex notation.
The convergence satisfies the Poisson equation $\Delta \psi = 2 \kappa$.
In the weak lensing regime ($\gamma \ll 1$), an elliptical object in the
source plane with complex ellipticity $e_s$ is mapped into an elliptical
image with a different ellipticity $e = e_s + \gamma$.
We refer the reader to the reviews of \citet{mellier99} and \citet{BS01}
for detailed accounts of weak lensing.

For a circularly symmetric lens, $\gamma$ is oriented tangentially
to the lens center and its amplitude at radius $r$ is 
$\gamma(r)=(\overline{\Sigma}(r)-\Sigma(r))/\scrit$. 
Since sources are randomly oriented, the tangential component of the observed
galaxies is an unbiased estimator of $\gamma$. When averaging the estimate
of $\gamma$ within an aperture of solid angle $\Omega$, containing
$N=\Omega n$ galaxies ($n$ is the number density of sources), the noise
dispersion of $\gamma$ is $\sigma_\gamma=\frac{\sigma_e}{\sqrt{N}}$ where
$\sigma_e\sim0.3$ is the intrinsic dispersion of source ellipticities
(along one component).

%------------------------------------
\subsection{Shear measurement}
In order to be consistent with our calculations of Sec.
\ref{subsec:projec-dens}, we define a pseudo-shear:
$\Gamma(\lambda)=(\overline{\Sigma}-\Sigma)/\rho_H r_{ta}$
and the corresponding noise level $\Gamma_N=\scrit\sigma_\gamma / \rho_H r_{ta}$.
For an EdS cosmology, and considering an annulus of inner and outer radii
$\lambda_1$ and $\lambda_2$ respectively, it is straightforward to write
$\Gamma_N$ in this useful form:
\begin{multline}
  \Gamma_N(\lambda_1,\lambda_2)= \, 2.16\, \frac{D_{os}}{D_{ls}} 
  \left(\frac{5\Mpc}{r_{ta}} \right)^2 (1+z_l)^3 \times \\
  \sqrt{\frac{30\, \mathrm{arcmin}^{-2}}{n}} 
  \left(\frac{\sigma_e}{0.3}\right) \frac{1}{\sqrt{\lambda_2^2-\lambda_1^2}}\;.
\end{multline}
This expression seems to indicate that the noise level will be lower for nearby
haloes. However, it hides the fact that low redshift haloes require a very
wide sky coverage for the outermost caustics ($\sim r_{\rm ta}/3$) to fit the
field of view of the observation. Consequently, intermediate redshift haloes
($z\sim0.2-0.5$) are the most interesting targets. In addition, nearby
(and thus large angular scale) clusters suffer from noise due to
large-scale structure (LSS) fluctuation integrated along the line-of-sight
and unrelated to the halo we are considering \citep{hoekstra03}.
For scales $\lesssim 15$ arcmin, the smearing of the shear
profile by LSS is minimised.

\begin{figure}[htbp]
  \centering
  \resizebox{\hsize}{!}{\includegraphics{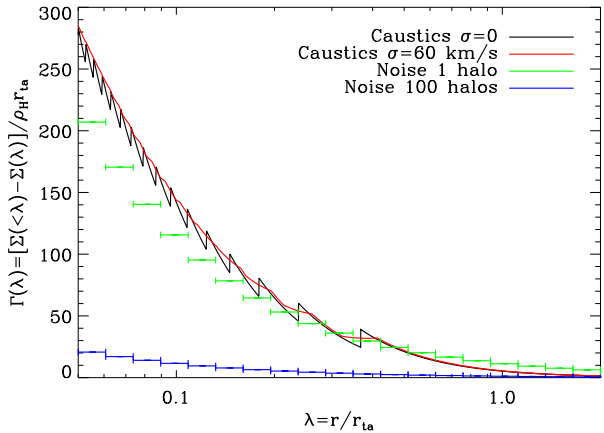}}
  \caption{\footnotesize $\Gamma(\lambda)$ contribution of caustics for two
    values of $\sigma$: cold medium $\sigma=0$ (black curve) and warm medium
    $\sigma\sim 60$km/s (red curve). The green (resp. blue) ``binned''
    curve is the noise level for one (resp. 100 stacked) halo(es).
  }
  \label{fig:Gamma}
\end{figure}

Figure \ref{fig:Gamma} shows $\Gamma$ as a function of $\lambda$ for the same
values of the thermal velocity dispersions $\sigma=0\kms$ and $\sigma=60\kms$.
Comparing these curves, one can see that the sawtooth patterns due to 
caustics survive significantly high temperatures.
Next, we consider the noise level for a fiducial halo at redshift $z_l=0.3$
and a turnaround radius $r_{ta}=5\Mpc$ which is a typical value for clusters
(upper green ``binned'' curve). With a single halo the detection of caustics
is impossible. If we are able to stack the signal from a few tens of
clusters (upper blue-binned curve), the noise level will be low enough to
be sensitive to caustics as a whole. However, we would detect a smooth
contribution which is indistinguishable from the halo itself.

Sawtooth patterns cannot be confused with the effect of substructures
since the contribution of the latter would be averaged out over the
azimuthal angle (most outer caustics have spherical symmetry) and once
rescaled, caustics always appear at the same place within haloes.
This is not the case for substructures which can appear at any radius
inside the host halo.

%...................
%...................

Consequently, the right way to probe the existence of caustics is to
measure the $\Gamma$ signal in excess/default relative to the extreme value
of $\sigma$, {\it e.g.}~ $\sigma\gtrsim 300 \kms$ as taken here. 
For this purpose, one needs $\gtrsim 100$ stacked clusters.
With the corresponding noise level, it would be possible to test the
thermal smoothing of caustics and put constraints on $\sigma$.
However, the sensitivity is poor and only upper limits
can be put on $\sigma$ with a realistic number of haloes.
For instance with $N=100$ (resp. 250) clusters, we could achieve
a limit $\sigma < 170\kms$ (resp. $40\kms$) at a $95.4\%$ confidence level.

When considering galaxies instead of clusters with turnaround radii 
about $10$ times smaller, the number of haloes required to achieve the same
detection level is $10^4$ times higher. So using a few $10^5$ galaxies
between $z\sim 0.1-0.5$ would yield the same results.

This required level of signal can easily be achieved 
with a wide and moderately deep survey like
the ongoing CFHTLS. Typically, a square-degree field of view will contain
a few such cluster-size haloes with $r_{\rm ta}\sim 5\Mpc$, a few thousand
elliptical galaxies with $r_{\rm ta}\sim 1\Mpc$ and tens of thousands of
spiral galaxies with $r_{\rm ta}\sim 500\kpc$. The total coverage of the
CFHTLS wide survey is 170 square degree and will contain a large enough
number of clusters/galaxies. Furthermore the wide fields of view are
well-suited to measure shear up to the outermost caustic of clusters
($1\Mpc=5\arcmin$ at $z=0.3$ and $4\arcmin$ at $z=0.5$). 

Spatial observations provide an improvement on shear measurements: (i)
the intrinsic dispersion in source ellipticities is lowered
$\sigma_e\sim0.2$ and (ii) the density of sources is increased
$n\sim50$ arcmin$^{-2}$. Hence the total number of haloes required to
achieve the same detection level can be lowered by a factor of
three. Future wide spatial surveys like SNAP or DUNE will provide
the required sky coverage.

%------------------------------------

\subsection{Halo-stacking issues}
So far, we have considered the observational difficulties encountered by
the dispersion in intrinsinc source ellipticities. However,
practical complications such as the difficult signal stacking process should
also be overcome before a detection of the caustics can be achieved.
To do so, high-precision measurements of the location of the center 
and the turnaround radius of
each halo are required. Otherwise, imperfect alignment/rescaling would
tend to blur the caustics spikes and reduce the sensitivity.
The center of a well-relaxed cluster coincides within one arc-second with the
center of the halo. Hence the only scaling factor is the turnaround radius
that can be related to the virial radius [see the sentence 
following expression (\ref{eq:rho3Dhalo})].

%...................
\begin{figure}[htbp]
  \centering
  \resizebox{\hsize}{!}{\includegraphics{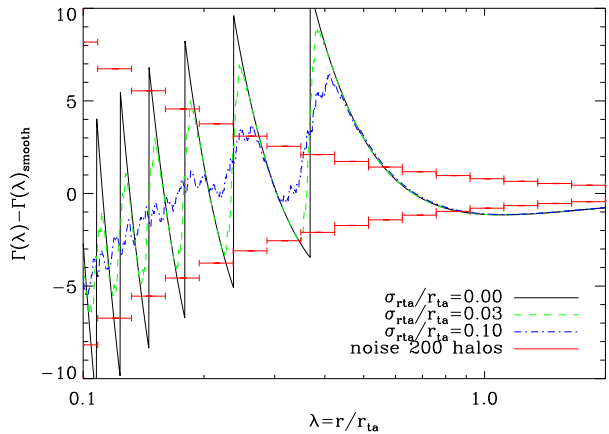}}
  \caption{\footnotesize Difference between the $\Gamma(\lambda)$ contribution
    of cold caustics and the smooth component. We illustrate the blurring
    effect of an imperfect knowledge of the turnaround radius of stacked
    haloes. The solid black line, dashed green line and dot-dashed
    blue line correspond to zero, 3\% and 10\% uncertainties, respectively. 
    As in \reffig{fig:Gamma}, the binned curve represents
    the noise level achieved with 200 stacked haloes. The convolution effect
    of error in the assumed/measured value of $r_{\rm ta}$ is important and a
    significant detection of caustics requires well-determined turnaround
    radii ($\lesssim$ a few percent relative accuracy). The very small scale
    oscillations in the plot are numerical artifacts and ideally the only
    sawtooth patterns are those due to caustics.
  }
  \label{fig:blur}
\end{figure}
%...................

Figure \ref{fig:blur} shows the blurring of caustics due to an imperfect
knowledge of the turnaround radius of each stacked halo. We consider a
dispersion around the true value of 3\% and 10\%. To properly detect
caustics, one needs a precise estimate of $r_{\rm ta}$. 
Provided dark matter is cold enough, $\sigma\lesssim$ a few $\kms$,
which is a realistic prescription, a reasonable number of stacked
haloes, $\sim 200$, can overcome the loss of a few percents of relative
precision caused by the errors in the determination of the turnaround radius.

Consequently the number of haloes necessary for detection needs to be
increased in order to achieve the necessary level of signal-to-noise
ratio. Moreover, additional external information for the
determination of $r_{\rm ta}$ such as X-rays or dynamical observations
in addition to weak lensing data are indispensable for a convincing
detection of caustics. The present-day state-of-the-art X-ray estimates 
for the scale radius is $\lesssim 10\%$ for nearby clusters ($z<0.2$)
\citep{arnaud05,pointecouteau05} and it seems that a similar precision
can be reached at higher redshifts ($0.4<z<0.7$) \citep{kotov05}.
In addition shear, which is used for caustic detection, also provides
constraints on the halo density profile. The virial radius of 
some clusters presenting strong and weak lensing features can be measured
with good accuracy ($\Delta r_{200}/r_{200} \lesssim 3$\%)
\citep{broadhurst05b,gavazzi05}. Consequently, future large 
cluster samples with X-rays and lensing data of present-day quality
will provide us with the necessary precision to probe dark matter caustics.

%======================================================================
\section{Discussion \& conclusion}\label{sec:conclu}
Although the Liouville theorem claims that singularities will survive,
they are likely to develop a complex topology in the course of evolution of
structures under self-gravity. It is not clear to what degree the merging
processes and substructures will smear out the caustics or complicate their
geometries. Here, we have considered the outer caustics of cluster-size
haloes which are expected to have suffered far less from mergers and due to
their relatively large separations are unlikely to have been washed away
by the dispersive-like effect of the substructures. These
caustics are the ones that contribute most to the lensing signal for
the following reasons. Their amplitude relative to the background smooth
halo component is more important as compared to the inner caustics.
The inner caustics smear most from thermalisation and imperfect stacking
and finally, the noise level increases toward the center of haloes thus
giving more weight to the outer caustics.
Hence, we have focused on caustics of cluster-size haloes and have argued
that they can be reasonably approximated by a selfsimilar spherical
accretion model, though the triaxiality of haloes is well established in
numerical simulations \citep{jing02tri}. However, caustic patterns should
exist in triaxial matter distributions. The singularities will have the
same elliptical symmetry and may be properly stacked from one halo to another
by choosing a subsample of apparently circular projected haloes or by
using the shear azimuthal variations to constrain the halo ellipticity.
We expect this effect to be comparable to the uncertainty in the halo scaling
radius (either turnaround or virial).

We have shown that the existence of dark matter caustics could be probed
by properly stacking the weak lensing signal of a reasonable
number of haloes. The main observational limitation is perhaps the
precise estimation of the turnaround radius, $r_{\rm ta}$, of superimposed
haloes but we have shown that the loss of a few percent relative accuracy
in the determination of $r_{\rm ta}$ (or asphericity) can be compensated for
by stacking about $200$ haloes.

Although the sensitivity is low, a detection of caustics provides an upper
bound for the temperature of dark matter, thus excluding hot dark matter
models. The sensitivity is not sufficient to distinguish between various 
cold dark matter candidates (like axions or neutralinos) since for 
most of the corresponding velocity dispersions, the shear
signal would be similar.\\
However, a detection of caustics would be a strong argument for the existence
of cold dark matter since alternative models like {\small MOND} could not
explain such density singularities and at most could serve
in place of a smooth halo (namely provide an equivalent effective
gravitational potential).\\
Putting constraints on the velocity dispersion of dark matter is a
challenging topic in modern cosmology since it offers
the possibility to pin-down an actual physical parameter of dark matter.
In this paper, we investigated the possibility of such a
measurement with the weak lensing effect of dark matter caustics.
The implicit observational hypothesis is that we can have 
a selfsimilar geometrical description of the caustic shell distribution
which depends  on a single characteristic scaling
parameter: {\it i.e.}~ the virial radius.

Wide field surveys such as the ongoing CFHTLS accompanied by X-ray observations
can provide the required statistics for a successful detection of caustics.
The number of haloes required to be superimposed will be lowered by a
further factor of 3 for future space-based experiments like SNAP or DUNE.
 
Here, we have used the first and most common caustic singularity,
that for which the density profile falls with inverse square-root
of the distance from the caustic. Caustics of higher-order singularities
can also appear in collisionless media and have already been
classified \citep{arnold86}. It remains a challenging task to 
generalize our work to haloes with non-vanishing eccentricity
where higher-rank caustics would occur and to examine if they 
can modify the magnification properties of the lensed images and account 
for anomalous image flux ratios.

\begin{acknowledgements}
Special thanks go to Sergei Shandarin for an ongoing collaboration
on dark matter caustics. We also thank Francis Bernardeau, Ed Bertschinger,
Monique Arnaud, Gary Mamon and Etienne Pointecouteau for many helpful
discussions and comments. R.G. is supported at Toulouse from
a CNRS postdoc contract \#1019 and at Oxford by a grant from
the Leverhulme Trust. R.M. is supported by European Gravitational
Observatory grant at the school of astronomy, university of Cardiff, UK. 
\end{acknowledgements}

\begin{scriptsize}
%\begin{small}
  \bibliographystyle{aa}
  \bibliography{references}
%\end{small}
\end{scriptsize}

\end{document}